\documentclass[aps,prl,twocolumn,superscriptaddress,nofootinbib,floatfix,longbibliography]{revtex4-2}
\usepackage{amsmath,amssymb}
\usepackage{graphicx}
\usepackage{xcolor}
\usepackage[colorlinks=true,citecolor=teal,linkcolor=teal,urlcolor=teal]{hyperref}
\usepackage{orcidlink}

\newif\ifdraft \drafttrue
\ifdraft

  \newcommand{\NK}[1]{{\color{teal}[NK: #1]}}
\else

  \newcommand{\NK}[1]{}
\fi

\newcommand{\Dthree}{D_{3000}^{\rm pkSZ}}
\newcommand{\tauthirty}{\tau_{\ell\geq30}}
\newcommand{\uK}{\mu{\rm K}^2}
\newcommand{\xHII}{\bar{x}_{\rm HII}}
\newcommand{\cMpc}{\,{\rm cMpc}}

\newcommand{\HII}{H\,{\sc ii}}

\begin{document}

\title{Transient early reionization allows for a high optical depth while \\ remaining consistent with kSZ measurements}
%

\author{Prakhar Bansal\,\orcidlink{0009-0000-7309-4341}}
\email{prakharb@umich.edu}
\author{Nicholas S. Kern\,\orcidlink{0000-0002-8211-1892}}
\email{nkern@umich.edu}
\author{Dragan Huterer\,\orcidlink{0000-0001-6558-0112}}
\email{huterer@umich.edu}
\affiliation{Department of Physics and Leinweber Institute for Theoretical Physics, University of Michigan,
450 Church St, Ann Arbor, MI 48109, USA}
\date{\today}

\begin{abstract}
Recent works have shown that a CMB optical depth of $\tau\approx0.09$, above the
Planck low-$\ell$ polarization value of 0.054, would weaken the preference of DESI data for dynamical dark energy and for a low neutrino mass. Apart from the low-$\ell$ CMB, another probe that prefers a $\tau\sim0.055$ is the kinetic
Sunyaev--Zel'dovich (kSZ) effect when combined with the end of hydrogen reionization as measured by the Ly$\alpha$ forest. In this work we use a new self-consistent model of supermassive Population~III stars, which partially ionize the intergalactic medium at high redshifts ($z\approx20$--$25$), to predict the kSZ signal. We show that such an early, transient reionization phase can raise $\tau$ to
$\approx0.09$ while remaining consistent with current kSZ and Ly-$\alpha$ forest constraints. We forecast that the Simons Observatory, combined with Ly-$\alpha$ forest and UV luminosity function data, can constrain the optical depth accumulated above
$z=12$ to $\pm0.008$ in our fiducial supermassive Pop~III scenario.
\end{abstract}


\maketitle

\paragraph*{Introduction.}
The Thomson-scattering optical depth to the cosmic microwave background (CMB),
$\tau$, measures the integrated column of free electrons along the line of
sight and is therefore sensitive to the timing and extent of reionization. Its
most precise determination, $\tau=0.054\pm0.007$, comes from the large-scale
(low-$\ell$) $E$-mode polarization of Planck~\cite{Planck:2018vyg}. This
measurement is limited by systematic effects that are difficult to
model~\cite{Planck:2016kqe,Pagano:2019tci,deBelsunce:2021mec}, and the optical
depth inferred without large-scale polarization, from the small-scale CMB
spectra and lensing, is increasingly in tension with
it~\cite{Giare:2023ejv,Sailer:2025lxj,Jhaveri:2025neg}. A higher $\tau$ would
also weaken the preference for dynamical dark energy and for a low neutrino
mass in DESI data~\cite{Sailer:2025lxj,Jhaveri:2025neg}. This motivates probes
of $\tau$ that are independent of low-$\ell$ polarization.

One such probe is the patchy kinetic Sunyaev--Zel'dovich (pkSZ) effect, which,
like low-$\ell$ $EE$, currently favors a low $\tau$. The kSZ effect is the
Doppler shift imparted on CMB photons that scatter off free electrons in bulk
motion~\cite{Sunyaev:1980vz,Ostriker:1986fua}. It is split into the
homogeneous kSZ (hkSZ) from the ionized intergalactic medium at $z\lesssim5.5$
and the pkSZ from reionization at $z\gtrsim5.5$~\cite{Park:2013mv}, and,
following standard practice, we call \emph{all} kSZ power from $z>5.5$ patchy.
Because the pkSZ traces the ionized regions during reionization, it constrains
$\tau$ independently of large-scale polarization.

At $\ell=3000$, where its amplitude is conventionally quoted, the pkSZ must be
separated from the thermal SZ effect, the cosmic infrared background, the
lensed CMB and the hkSZ. The South Pole Telescope (SPT) has measured the total
kSZ power and set upper limits on the pkSZ~\cite{SPT:2020psp,SPT-3G:2026mcr},
and the Simons Observatory (SO)~\cite{SimonsObservatory:2018koc} is expected to
make the pkSZ a powerful probe of reionization.

Together with the Ly$\alpha$ forest, which places the end of reionization at
$5.3\lesssim z\lesssim6$~\cite{Bosman:2021oom}, the pkSZ disfavors galaxy-driven
histories extended enough to reach $\tau\approx0.09$ at
$\approx2\sigma$~\cite{Cain:2025usc}. However, the pkSZ at $\ell=3000$ is
dominated by ionized regions of size $R\sim20\,{\rm Mpc}$ and is comparatively
insensitive to much smaller bubbles~\cite{Aggarwal:2026ogm}. If such bubbles
carry an appreciable fraction of $\tau$ while leaving the pkSZ at $\ell=3000$
largely unaffected, a pkSZ-based inference would underestimate $\tau$.

A physically motivated realization of
this possibility is a period of transient reionization, in which a population of supermassive population III
stars form at high redshifts ($z\sim25$) and partially reionize the intergalactic medium (IGM) before it recombines, leading to an additional source of free electrons along the line-of-sight.
One widely debated scenario comes from a weakly interacting massive particle (WIMP) dark matter candidate that undergoes self-annihilation.
In the cores of high-redshift dark matter halos, self-annihilation feedback can in theory become large enough to prevent protostellar collapse of standard population III stars, and enable a long ``dark star'' accretion phase that may in theory reach supermassive states $\sim10^5\ M_\odot$.
First studied in Refs.~\cite{Spolyar:2007qv,Freese:2008wh,Natarajan2009}, the physics of these objects has long been debated and studied by subsequent works \citep{Stacy2014, Freese:2015mta, Banik:2016qww, Singh:2023beo, Cammelli:2024uhh, Tan:2024ysr, Tan:2025cua, Tan:2025obi, Upadhyay:2026jks}.
In this work we are concerned with the ionizing radiation from these sources after they collapse from their dark star phase and reach the zero-age main sequence (ZAMS), and as a consequence we will refer to them as (supermassive) Pop~III.1 stars,\footnote{We treat all Pop~III.1 stars as supermassive, with the understanding that a form of radiative support such as annihilating dark matter is needed for this.} which form only in isolated, pristine minihalos \citep{McKee2008}.
Aggarwal et al.~\cite{Aggarwal:2026ogm} investigated the impact of such flash-ionizing sources on the pkSZ signal using the \textsc{AMBER} simulations;
however, as the authors acknowledge, their flash \HII{} regions were painted
onto a separately simulated reionization field, so that they neither recombine
nor interact with the galaxy-driven ionization fronts.

In this work, we revisit this question using the \textsc{21cmFAST}
code~\cite{Mesinger:2010ne,Park:2018ljd,Qin2020,Murray:2020trn,Davies:2025wsa} as modified in
Ref.~\cite{Kern:inprep}, which models Pop~III.1-driven feedback self-consistently. We compute the resulting pkSZ signal and assess whether current measurements could rule out a population of ionized bubbles at high redshifts that would contribute significantly to the CMB optical depth. We further show that the pkSZ, like the
optical depth inferred from CMB spectra at $\ell\geq30$, measures mostly the
total optical depth, so that an early component can be isolated only once the
galaxy-driven history is anchored: by the Ly$\alpha$ forest, which fixes its
end, and by UV luminosity functions, which fix how drawn-out it is. Measuring
the pkSZ in several multipole bins then helps separate early from late
ionization. Below we describe the model, show how much these sources contribute to $\tau$ and why they are hidden from current kSZ constraints, and then forecast how future CMB surveys such as SO can begin to constrain them.
Note that the goal of this work is not to weigh in on the feasibility of dark star or supermassive population III star formation. Rather, we are interested in asking the question: if these objects exist as prescribed in the literature, what impact do they have on current and future observables?

\paragraph*{Model.}
We refer to \citet{Kern:inprep} for details of the semi-numerical implementation of Pop~III.1 physics.
The simulations use \textsc{21cmFAST}~\cite{Davies:2025wsa} in a $252\cMpc$ box with
$504^3$ ($0.5\cMpc$) ionization cells and $1008^3$ initial conditions, from
$z=40$ to $5.5$, with inhomogeneous recombinations and spin-temperature
fluctuations.
We use standard 21cmFAST treatment of atomic-cooling galaxies and minihalos.
Our fiducial galaxy-only model reaches $\bar x_{\rm HII}=0.9$ at $z_{90}=6.04$ and
$0.95$ at $5.90$, consistent with the Ly$\alpha$ forest and quasar
damping wings (Supplemental Material), and gives $\tau=0.060$. We compute $\tau$ from
the mass-weighted ionized fraction, because the free-electron density follows
the gas, $n_e\propto x_{\rm HII}(1+\delta)$, and ionized regions are overdense during reionization. 
We include helium reionization as in the Planck analysis, singly ionized with
hydrogen and doubly ionized at $z=3$, so that
\begin{equation}
  \tau(z)=\sigma_Tc\int_0^z\frac{(1+z')^2}{H(z')}\,\bar n_{e,0}(z')\,
  \langle x_{\rm HII}(1+\delta)\rangle_{z'}\,dz',
  \label{eq:tau}
\end{equation}
where $\bar n_{e,0}=\bar n_{{\rm H},0}+\bar n_{{\rm He},0}$, plus a further
$\bar n_{{\rm He},0}$ at $z'<3$, is the comoving electron density of fully
ionized gas.

The supermassive Pop~III.1 module~\cite{Kern:inprep} places sources in cells at a rate set by the growth of the collapsed fraction above a minimum halo mass, forbids a new source within an isolation distance $d_{\rm iso}$ of an existing one,
and gives each source an ionizing photon rate $Q_{\rm ion}$ for a lifetime
$t_{\rm life}$, producing a relic \HII{} region of radius $R_{\rm HII}$ [Supplemental
Material, Eq.~(\ref{eq:RHII})] that cools and recombines over time.


\emph{kSZ.} Along the line of sight the kSZ temperature shift is
$\Delta T/T=-\sigma_T\int d\chi\,a\,n_e\,e^{-\tau}\,\hat{n}\cdot\mathbf{v}/c$.
We compute its angular power in the Limber approximation~\cite{Park:2013mv},
\begin{equation}
  C_\ell=\frac{(\sigma_T\bar n_{e,0})^2}{2}\int\!\frac{c\,dz}{H(z)}\,
  \frac{(1+z)^4e^{-2\tau(z)}}{\chi^2}\,P_{q_\perp}\!\left(\frac{\ell}{\chi},z\right),
  \label{eq:limber}
\end{equation}
where $P_{q_\perp}$ is the power of the transverse momentum field
$\mathbf{q}=(1+\delta_e)\mathbf{v}/c$. We build it from the electron-density
auto-spectrum and the electron--matter cross-spectrum measured in the
simulation boxes, with linear velocities. We include the density--velocity
cross term of Ref.~\cite{Park:2013mv} and correct the measured spectra for the
smoothing of the simulation grid. Evaluating
$P_{q_\perp}$ directly in three dimensions avoids replicating the box along
the line of sight. The pkSZ is the $z>5.5$ part of this integral. Across our four
realizations $\Dthree$ scatters by 0.7\% and the flash increment, from
paired runs, by 1\% (Supplemental Material).
Velocity modes larger than the box, included through linear theory, supply
$\approx40\%$ of $\Dthree$ but change the flash and no-flash spectra by the
same factor (Supplemental Material, Sec.~\ref{sec:ksz}).

\begin{figure*}[t]
\begin{minipage}[b]{0.485\textwidth}
  \includegraphics[width=\linewidth]{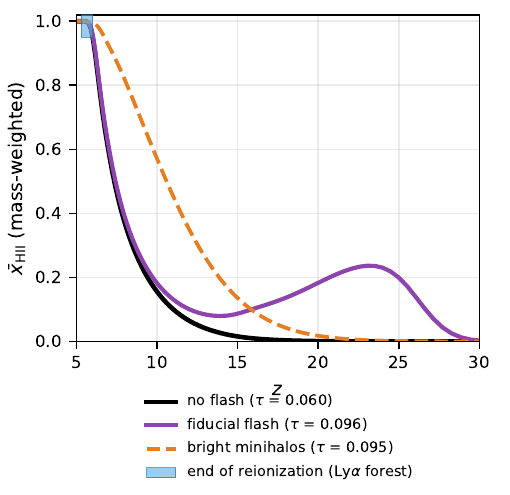}
\end{minipage}\hfill
\begin{minipage}[b]{0.485\textwidth}
  \includegraphics[width=\linewidth]{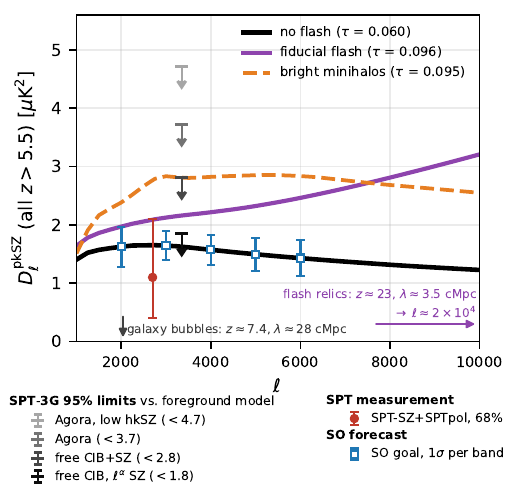}
\end{minipage}

\begin{minipage}[t]{0.485\textwidth}
  \caption{\textbf{A Pop~III.1 flash adds optical depth without moving the end
  of reionization.} Mass-weighted mean ionized fraction for the galaxies-only
  fiducial model (black, $\tau=0.060$) and with the reference
  Pop~III.1 model (purple, $\tau=0.096$), run from the same initial
  conditions and averaged over four realizations, and a bright-minihalo
  model that reaches a similar $\tau$ through an extended
  galaxy-driven history (orange dashed, $\tau=0.095$; one realization
  in a $150\cMpc$ box with $1\cMpc$ cells). Box: end of reionization from the Ly$\alpha$ forest,
  $5.3\lesssim z\lesssim6$~\cite{Bosman:2021oom}.}
  \label{fig:histories}
\end{minipage}\hfill
\begin{minipage}[t]{0.485\textwidth}
  \caption{\textbf{The Pop~III.1 kSZ power lives at small scales.} pkSZ power
  (all kSZ from $z>5.5$) for the three models of Fig.~\ref{fig:histories}. The
  bright-minihalo model is at lower resolution, which lowers its high-$\ell$
  power. Arrows at the bottom: the multipoles $\ell=k\chi$ at which
  galaxy-driven bubbles (at the reionization midpoint) and flash relics (at
  the flash peak) carry their ionization power. Grey-to-black arrows: SPT-3G
  $95\%$ upper limits on the pkSZ at $\ell=3000$ for four foreground models,
  drawn at $\ell=3350$ (Supplemental Material, Sec.~\ref{sec:ksz}). Red point:
  the SPT-SZ+SPTpol pkSZ measurement ($68\%$; at
  $\ell=2700$)~\cite{SPT:2020psp}. Blue squares: forecast SO $1\sigma$ errors
  on $\Delta\ell=1000$ bandpowers around the no-Pop~III.1 model.}
  \label{fig:pksz}
\end{minipage}
\end{figure*}

\paragraph*{A high optical depth from Pop III.1-driven transient reionization.}
Figure~\ref{fig:histories} shows the ionization history with and without
the Pop III.1 sources, run from the same initial conditions. The flash ionizes up to
23\% of the gas at $z_{\rm fl}\approx23$; the relics then
recombine to a plateau of $\approx8\%$ at $z\approx14$ and are absorbed
into the galaxy-driven fronts, leaving the late history essentially unchanged:
$z_{90}$ moves by 0.03 and the midpoint by 0.07, both well
within the forest constraints. The flash raises $\tau$ from 0.060 to
0.096; of the increment $\Delta\tau=0.037$, 95\% is
accumulated at $z>12$.

Figure~\ref{fig:pksz} is our main result: it shows the impact of Pop~III.1 sources on the pkSZ power spectrum. Without Pop~III.1, the spectrum peaks at $\ell\approx2750$ with
$\Dthree=1.64\,\uK$ and falls toward small scales. The new sources add
$0.48\,\uK$ at $\ell=3000$, leaving $\Dthree=2.13\,\uK$, below
three of the four SPT-3G upper limits~\cite{SPT-3G:2026mcr}. The tightest
limit, $1.8\,\uK$ (free CIB with a power-law tSZ), is exceeded, but the
no-Pop~III.1 model already nearly saturates it. The excess grows with $\ell$, to
$0.64\,\uK$ at $\ell=4000$ and a factor 2.6 over the
no-flash power at $\ell=10^4$: a 60\% increase in $\tau$ that the pkSZ at
$\ell=3000$ registers only as a 30\% rise in amplitude, while the spectral
\emph{shape} changes sign -- rising where galaxy-driven reionization falls.
For SO, with foregrounds and cosmology marginalized, the excess is
$2.0\sigma$ in the $\ell=3000$ bandpower but 2.5--3.3$\sigma$ in
each $\Delta\ell=1000$ band at $\ell=4000$--$6000$ (Fig.~\ref{fig:pksz}).
A galaxy-driven route to the same optical depth fares worse. Our
bright-minihalo model reaches $\tau=0.095$ through an extended history
(Fig.~\ref{fig:histories}) and adds $1.2\,\uK$ at $\ell=3000$
($\Dthree=2.83\,\uK$, at or above two of the four limits). This realization
also ends reionization early ($z_{90}\approx6.8$), but extended histories that
end on time are similarly costly in pkSZ~\cite{Cain:2025usc}.

\begin{figure}
  \includegraphics[width=\columnwidth]{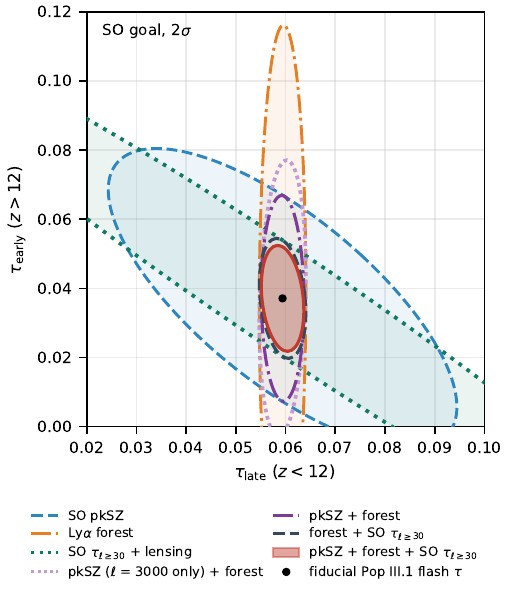}
  \caption{\textbf{Exposing the flash requires anchoring the late history.}
  Fisher forecast for SO around the fiducial flash universe, in the plane of
  the optical depth accumulated below and above $z=12$. Ellipses span the
  marginalized $\pm2\sigma$ errors. Light fills: each probe alone, the pkSZ in
  four bandpowers at $\ell=3000$--$6000$ (blue dashed), the Ly$\alpha$ forest
  (orange) and $\tauthirty$ (green dotted). Lines: combinations. Light purple
  dotted uses only the $\ell=3000$ bandpower. All contours include the UV-LF
  calibration of the galaxy model and the priors on the ten flash parameters,
  with $\Lambda$CDM and the late-time kSZ marginalized
  (End Matter, Table~\ref{tab:sigtau}). Dot: the fiducial flash,
  $(\tau_{\rm late},\tau_{\rm early})=(0.059,0.037)$.}
  \label{fig:tauplane}
\end{figure}

\paragraph*{Why the flash hides.}
The flash hides because its ionized regions are far smaller than the scales
the pkSZ at $\ell=3000$ probes. The kSZ at multipole $\ell$ samples ionization
fluctuations at comoving wavenumber $k\approx\ell/\chi$
[Eq.~(\ref{eq:limber})]: the bulk velocities are coherent over
$\sim100\cMpc$, so the small-scale structure comes from the electrons.
Galaxy-driven bubbles carry their ionization power at wavelengths of
$\approx28\cMpc$ at the reionization midpoint ($z\approx7.4$),
which maps to $\ell\approx2000$; the no-flash spectrum accordingly
peaks at $\ell\approx2750$ (Fig.~\ref{fig:pksz}). The flash's relics
carry theirs at $\approx3.5\cMpc$ at $z\approx23$, eight times
smaller, which maps to $\ell\approx2\times10^4$. At that epoch,
$\ell=3000$ probes wavelengths of $\approx23\cMpc$, far larger than the relics.

Figure~\ref{fig:pksz} shows exactly this. The flash adds
$0.48\,\uK$ at $\ell=3000$ ($\approx13\,\uK$ per unit $\tau$)
but $2.0\,\uK$ at $\ell=10^4$ ($\approx54\,\uK$ per unit
$\tau$), and its excess peaks at $\ell\approx1.7\times10^4$. Most of
the flash's kSZ power therefore lies above the multipoles at which the pkSZ is
currently measured.

The flash is also invisible to the duration of reionization that pkSZ
analyses infer. Constraints are usually quoted on
$\Delta z_{50}=z(\xHII=0.25)-z(\xHII=0.75)$ at a fixed midpoint
$z_{\rm re}$~\cite{SPT-3G:2026mcr}; a flash that never ionizes more than $25\%$ of
the volume cannot move either, and ours (peak $\xHII=0.23$) shifts
$\Delta z_{50}$ by only 0.15. A duration analysis would read its
$+0.48\,\uK$ as slightly longer reionization, never as an early
flash. The flash's signature is instead in the \emph{shape} of the
spectrum~\cite{Chen:2022lhr,Gorce:2020pcy,Jain:2023jpy}.

\paragraph*{How the information can be accessed.}
Having seen why the Pop~III.1 model hides kSZ signal at $\ell=3000$, we now
forecast whether the data available by the end of the decade can nevertheless
find it. We combine four data sets, each sensitive to a different part of the
reionization history (Supplemental Material, Sec.~\ref{sec:forecast}).
(i)~The pkSZ: the TT, EE and TE spectra of SO at goal
sensitivity~\cite{SimonsObservatory:2018koc} at $30\leq\ell\leq6900$ and its
lensing spectrum, compressed into bandpowers of width $\Delta\ell=1000$ at
$\ell=3000$--$6000$, with $\Lambda$CDM and the late-time ($z<5.5$) kSZ
marginalized. (ii)~$\tauthirty$, the optical depth from the same spectra
without large-scale polarization, with $\sigma_\tau=0.0086$ against $0.016$
from Planck~\cite{Giare:2023ejv}. We impose no external $\tau$. (iii)~The
Ly$\alpha$ forest: four quasar damping wings at
$z=7.0$--$7.5$~\cite{Davies:2018yfp,Wang:2020zae,Yang:2020J1007} and the end of
reionization, $x_{\rm HI}\lesssim5\%$ at
$5.3<z<5.8$~\cite{Cain:2025usc,Bosman:2021oom}. (iv)~The UV luminosity
functions (LFs), as priors of $0.30$\,dex on $F_{\star,10}$ and $0.07$ on the
slope $\alpha_\star$ of the star-formation efficiency~\cite{Park:2018ljd}, which
fix how drawn-out galaxy-driven reionization is. The model responses come from
the simulations, in these three galaxy parameters and the ten Pop~III.1
parameters of Ref.~\cite{Kern:inprep} with their priors.

We present the constraints in the plane of two partial optical depths:
$\tau_{\rm late}$, accumulated at $z<12$, and $\tau_{\rm early}$, accumulated
at $z>12$ (Fig.~\ref{fig:tauplane}). The fiducial flash universe has
$\tau_{\rm late}=0.059$ and $\tau_{\rm early}=0.037$.
Galaxies source almost all of $\tau_{\rm late}$, and the flash sources almost
all of $\tau_{\rm early}$: without the flash, $\tau_{\rm early}$ would be
0.002. The plane therefore separates the two sources of ionization,
and it also makes the problem posed in the Introduction visible. The pkSZ and
the forest together, for example, measure $\tau_{\rm late}$ to
$\pm0.0021$, mostly through the forest and the UV LFs, but constrain
$\tau_{\rm early}$ only to $\pm0.015$. Most kSZ analyses, whether
they forward-model the galaxy population or parametrize the history with a
$\tanh$ form, assume that early ionization is negligible. This assumption
restricts the posterior to the slice $\tau_{\rm early}=0$ and has two
consequences. First, the error on the total optical depth is underestimated.
For the pkSZ with the forest it falls from $\pm0.015$ to
$\pm0.0021$, because the data then measure only $\tau_{\rm late}$.
Second, the inferred total optical depth is biased low. It equals the
galaxy-driven value, 0.060, rather than the true 0.096.

Once early ionization is allowed, no single probe finds the flash
(Fig.~\ref{fig:tauplane} and Table~\ref{tab:sigtau} in the End Matter). The pkSZ alone measures a combination close to the
total optical depth: its contour runs along lines of constant $\tau$, and it
leaves $\sigma(\tau_{\rm early})=0.022$, comparable to the flash's
$\Delta\tau_{\rm early}=0.035$. $\tauthirty$ measures the total
directly and is degenerate in nearly the same direction; together the two
give $\sigma(\tau_{\rm early})=0.019$. The forest is orthogonal to both: the damping
wings, with the UV-LF calibration of $\alpha_\star$, pin
$\tau_{\rm late}$ to $\pm0.0023$ but say nothing about $z>12$.
Anchoring the late history breaks the degeneracy. With $\tau_{\rm late}$
fixed, $\tauthirty$ measures $\tau_{\rm early}\simeq\tauthirty-\tau_{\rm
late}$, and the pkSZ measures it through the shape of its spectrum, where the
flash's power rises with $\ell$ while the galaxy-driven power falls
(Fig.~\ref{fig:pksz}). The forest with $\tauthirty$ gives
$\sigma(\tau_{\rm early})=0.0087$, the forest with the pkSZ
0.015, and all three 0.0077, a fifth of the fiducial
flash's early optical depth. The same data measure the
galaxy-driven optical depth even with the flash free, $\sigma(\tau_{\rm
late})=0.0021$: an order of magnitude below the $\approx0.03$ that a
galaxy-driven route to $\tau\approx0.09$ would add (Fig.~\ref{fig:histories}).

Note that current analyses quote the reionization kSZ as a
single number: SPT-3G reports the kSZ power in several multipole bins, but the
patchy constraint and the reionization inferences built on it use $\ell=3000$
alone~\cite{SPT-3G:2026mcr}. With the forest, the $\ell=3000$ bandpower gives
$\sigma(\tau_{\rm early})=0.020$, and the four bandpowers give
0.015 (light against dark purple in Fig.~\ref{fig:tauplane}). 
Reporting the
pkSZ in several multipole bins is therefore a concrete and inexpensive step for
future analyses: it turns a measurement of the duration of reionization into
one of when the optical depth was accumulated.
Finally, $\tauthirty$ and the shape of the pkSZ spectrum are two independent
routes to $\tau_{\rm early}$ that rest on different assumptions. $\tauthirty$
relies on lensing to break the $A_se^{-2\tau}$ degeneracy within
$\Lambda$CDM, whereas the pkSZ relies on the simulated relation between the
reionization history and the shape of the high-$\ell$ kSZ spectrum. The two
are best used as cross-checks.

\paragraph*{Discussion.}
\emph{Comparison with Ref.~\cite{Aggarwal:2026ogm}.} Ref.~\cite{Aggarwal:2026ogm}
painted Pop~III.1 bubbles seeded on density peaks onto a separate reionization
model and found that a flash with $\Delta\tau\approx0.03$ adds
$\approx1.9\,\uK$ at $\ell=3000$, about five times more per
unit $\tau$ than our self-consistent flash. This difference likely lies in the assumed
large-scale clustering of the relics. In our simulations the relics lose the
bias of their host halos by the flash peak ($b\approx0.7$ against
$\approx7.7$ at $z\approx23$), and relics that kept the host bias
would have enough additional power on the scales probed at $\ell=3000$ to
account for the difference (Supplemental Material, Sec.~\ref{sec:bias}). 

\emph{Galaxy-driven alternatives.} With SO, the forest and the UV LFs, the
galaxy-driven optical depth is measured to $\pm0.002$ even when a flash is
allowed, so within our galaxy model a high $\tau$ must be accumulated early.
Calibrated by the UV LFs, the \textsc{21cmFAST} parametrization cannot produce
the very extended histories of Ref.~\cite{Cain:2025usc}, which begin
reionization earlier and raise $\tau$ while keeping its end fixed. The forest
and $\tauthirty$ cannot distinguish such histories from a flash, but the pkSZ
can, because an extended galaxy-driven history raises the power at all
multipoles (the bright-minihalo model in Fig.~\ref{fig:pksz}), whereas the
flash adds power mainly at high $\ell$. Our forecast therefore understates the
role of the pkSZ relative to $\tauthirty$.

\emph{Current data.} The optical depth inferred without large-scale
polarization is currently high: $\tauthirty=0.078\pm0.016$ from Planck with
lensing~\cite{Giare:2023ejv}, and $0.090\pm0.012$ with DESI
BAO~\cite{Sailer:2025lxj}, compared with $0.054\pm0.007$ from large-scale
polarization~\cite{Planck:2018vyg}. For $\tau_{\rm late}\approx0.06$, these
values correspond to $\tau_{\rm early}\approx0.02$--$0.03$, the range a
Pop~III.1 flash supplies. Applying our framework to the forest and Planck's
$\tauthirty$ gives $\tau_{\rm early}=0.023\pm0.015$ and
$\tau_{\rm late}=0.058\pm0.002$. Adding the SPT-3G total kSZ at
$\ell=3000$~\cite{SPT-3G:2026mcr} barely tightens this ($\sigma=0.014$) and
moves $\tau_{\rm early}$ to 0.016--0.026, depending on the foreground model
(Supplemental Material, Sec.~\ref{sec:current}). Current data therefore allow
both a flash like ours and none, and SO would halve the error on
$\tau_{\rm early}$. A more realistic analysis would fit the reionization
parameters to the forest, the Planck $\tauthirty$ and the SPT pkSZ jointly
with a full likelihood, rather than combine linearized Fisher constraints.

\emph{Caveats.} Our pkSZ uses linear velocities in the Limber approximation
and neglects the non-Gaussian four-point contribution to the momentum power,
which may raise $\Dthree$ by 10--20\%~\cite{Alvarez:2015xzu,Lopez:2025bpg}.
The late-time kSZ below $z=5.5$ is modelled with a template rather than
simulated. The flash relics span only a few cells, so the high-$\ell$ shape is
less robust than the power at $\ell=3000$--$4000$ (Supplemental Material). 
We treat the foreground power left by the ILC as known, as is common in previous forecasting studies.
Marginalizing a $10\%$ uncertainty on its amplitude changes $\sigma(\tau_{\rm early})$ by $\lesssim4\%$.

\emph{Conclusions.} A self-consistently simulated Pop~III.1 transient reionization era can raise
$\tau$ to $\approx0.1$ without moving the end of reionization.
This model is not in tension with any current pkSZ limits
because its small, weakly
biased relics place their kSZ power at $\ell\gtrsim10^4$. The pkSZ, the
Ly$\alpha$ forest and the UV luminosity functions nevertheless measure the
galaxy-driven optical depth precisely, so an excess in $\tau$ would have to
come from early ionization. Neither the pkSZ nor $\tauthirty$, the optical
depth from CMB spectra at $\ell\geq30$, isolates this excess on its own,
because both are sensitive mainly to the total optical depth. Anchoring the
late history with the forest breaks this degeneracy. Together with the
forest, the pkSZ and $\tauthirty$ from SO would measure $\tau(z>12)$ to
$\pm0.008$ around our fiducial flash without large-scale
polarization. 
Extending pkSZ measurements to higher multipoles allows one to probe more complex reionization histories.
These Pop~III.1 sources also leave a distinct 21-cm signal at $z\approx20$--$25$, studied in Ref.~\cite{Kern:inprep}.
Future work will investigate how to combine auto and cross-correlation measurements of the pkSZ signal with other high-redshift probes to constrain the early contribution to $\tau$.

\begin{acknowledgments}
We thank Simone Ferraro, Kendrick Smith, Feige Wang and Xiangyu Jin for useful discussions.
This work has been supported by the Department of Energy under contract
DE-SC0019193.
This research was supported in part through computational resources and services provided by Advanced Research Computing at the University of Michigan, Ann Arbor.
We gratefully acknowledge support from the Leinweber Institute for Theoretical Physics at the University of Michigan.

AI tools were used to generate some of the code and unit-tests for this work, in addition to outlining and proofreading some of the manuscript text.

\end{acknowledgments}

\bibliography{refs}

\onecolumngrid
\vspace{2ex}
\noindent\begin{minipage}{\textwidth}
\begin{center}
  {\large\bfseries End Matter}
\end{center}
\makeatletter\def\@captype{table}\makeatother
  \caption{\label{tab:sigtau} Forecasts for SO (goal sensitivity) around the
  fiducial flash universe: $1\sigma$ errors on $\tau_{\rm late}$ and
  $\tau_{\rm early}$ with the ten flash parameters free (the fiducial flash has
  $\Delta\tau_{\rm early}=0.035$ above the galaxies-only value). Every combination
  includes the UV-LF calibration of the galaxy model, except in the
  ``without UV LF'' columns, where the star-formation slope $\alpha_\star$ is
  free, and, for the flash universe, the priors on the flash parameters. ``pkSZ'': the four
  bandpowers at $\ell=3000$--$6000$. ``--'': unconstrained ($\sigma>0.1$).}
  \begin{ruledtabular}
  \begin{tabular}{lcccc}
  & \multicolumn{2}{c}{with UV LF} & \multicolumn{2}{c}{without UV LF} \\
  data & $\sigma(\tau_{\rm late})$ & $\sigma(\tau_{\rm early})$ & $\sigma(\tau_{\rm late})$ & $\sigma(\tau_{\rm early})$ \\
  \hline
  pkSZ $\ell=3000$                         & 0.055  & 0.037  & -- & 0.044 \\
  pkSZ                                     & 0.018  & 0.022  & 0.024 & 0.029 \\
  forest                                   & 0.0023 & 0.040  & 0.0049 & 0.040 \\
  $\tauthirty$                             & 0.038  & 0.038  & -- & -- \\
  pkSZ + $\tauthirty$                      & 0.017  & 0.019  & 0.019 & 0.019 \\
  pkSZ $\ell=3000$ + forest                & 0.0023 & 0.020  & 0.0047 & 0.023 \\
  pkSZ + forest                            & 0.0021 & 0.015  & 0.0038 & 0.021 \\
  forest + $\tauthirty$                    & 0.0023 & 0.0087 & 0.0049 & 0.0096 \\
  pkSZ $\ell=3000$ + forest + $\tauthirty$ & 0.0023 & 0.0079 & 0.0047 & 0.0094 \\
  pkSZ + forest + $\tauthirty$             & 0.0021 & 0.0077 & 0.0035 & 0.0092 \\
  \hline
  today: Planck $\tauthirty$ + forest      & 0.0023 & 0.015 & 0.0049 & 0.015 \\
  \end{tabular}
  \end{ruledtabular}
\end{minipage}

\clearpage
\onecolumngrid
\begin{center}
  {\large\bfseries Supplemental Material}
\end{center}
\setcounter{section}{0}\setcounter{figure}{0}\setcounter{table}{0}\setcounter{equation}{0}
\setcounter{secnumdepth}{2}
\renewcommand{\thesection}{S\Roman{section}}
\renewcommand{\thefigure}{S\arabic{figure}}
\renewcommand{\thetable}{S\arabic{table}}
\renewcommand{\theequation}{S\arabic{equation}}

\section{Simulations}
\label{sec:sims}

\emph{Code and setup.} We use \textsc{21cmFAST}~\cite{Davies:2025wsa} with the
Pop~III.1 flash module of Ref.~\cite{Kern:inprep}. All runs use a $252\cMpc$ box
with $504^3$ Eulerian cells ($0.5\cMpc$) and initial conditions on a $1008^3$ grid,
second-order Lagrangian perturbation theory, inhomogeneous recombinations,
spin-temperature fluctuations, relative baryon--dark-matter streaming velocities, $\Delta\ln(1+z)=0.02$ steps from
$z=40$ to $5.5$. See \citet{Kern:inprep} for more details.
We adopt a Planck cosmology $\Lambda$CDM with $h=0.6766$,
$\Omega_m=0.3096$, $\Omega_b=0.0490$, $n_s=0.9665$ and $\sigma_8=0.810$ \citep{Planck:2018vyg}.
Every Pop~III.1 run is paired with a standard 21cmFAST run without Pop~III.1 physics with identical initial conditions.
We run 4 independent seeds for the initial conditions, propagating this to the Pop~III.1 source placement and predicted pkSZ signal, finding negligible Monte-Carlo noise with 4 seeds.
Atomic-cooling galaxies and minihalos are treated in the standard 21cmFAST manner, tracking their ionization, Lyman-Werner, and X-ray output.


\emph{Supermassive Pop~III.1 flash.} Sources are placed in cells at a rate set by the growth of the collapsed fraction above $M_{\rm min}$, with the mean rate normalized to the mass function of Ref.~\cite{Watson:2012mt}. They ignite after a delay
$t_{\rm acc}$ and shine with ionizing rate $Q_{\rm ion}$ for $t_{\rm life}$. A new source is forbidden within an isolation distance $d_{\rm iso}=a_{\rm iso}R_{\rm HII}$ of an existing one~\cite{Banik:2016qww}, and in gas whose total electron fraction exceeds $\approx10^{-3}$, where
molecular cooling would form Pop~III.2 stars instead. Each source produces a relic \HII{} region of comoving radius
\begin{equation}
  R_{\rm HII}\simeq1.10\cMpc
  \left(\frac{Q_{\rm ion}}{10^{53}\,{\rm s^{-1}}}\right)^{1/3}
  \left(\frac{t_{\rm life}}{10\,{\rm Myr}}\right)^{1/3}.
  \label{eq:RHII}
\end{equation} Relics recombine with clumping factor $C_{\rm HII}$ and are absorbed into the galaxy-driven ionization field.
Pop~III.2 stars form in molecular-cooling minihalos only in gas previously processed by a Pop~III.1 source, within its isolation distance.

\begin{table}[h]
\caption{Parameters of the reference models (Figs.~\ref{fig:histories}
and~\ref{fig:pksz}) and of the Fisher forecast: fiducial value and Gaussian
prior width (``fixed'': not varied in the forecast). The no-flash model uses the
same galaxy and minihalo values with the flash switched off. Logarithms are
base 10. See Ref.~\cite{Kern:inprep} for the flash parameters and
Ref.~\cite{Davies:2025wsa} for the galaxy and minihalo parameters.}
\label{tab:params}
\begin{ruledtabular}
\begin{tabular}{lllc}
 & parameter & fiducial & prior \\
\hline
Galaxies & $\log F_{\star,10}$ & $-1.25$ & $0.30$\,dex \\
 & $\log F_{\rm esc,10}$ & $-1.32$ & free \\
 & $\alpha_\star$ & $0.6$ & $0.07$ \\
 & $t_\star$, $\alpha_{\rm esc}$ & $0.5$, $0$ & fixed \\
 & $\log[L_X/{\rm SFR}/({\rm erg\,s^{-1}}\,M_\odot^{-1}\,{\rm yr})]$, $E_0$ & $40.5$, $500$\,eV & fixed \\
\hline
Minihalos & $\log F_{\star,7}$, $\log F_{\rm esc,7}$, $\log L_{X,\rm mini}/{\rm SFR}$ & $-1.75$, $-2.9$, $41.0$ & fixed \\
 & LW and streaming feedback $(A_{\rm LW},\beta_{\rm LW},A_{\rm vcb},\beta_{\rm vcb})$ & $(2.0,0.6,1.0,1.8)$ & fixed \\
\hline
Flash & isolation factor $a_{\rm iso}$ & $1.40$ & $0.5$ \\
 & $\log(t_{\rm life}/{\rm Myr})$ & $0.90$ & $0.45$\,dex \\
 & accretion delay $t_{\rm acc}$ [Myr] & $25$ & $75$ \\
 & $\log(Q_{\rm ion}/{\rm s^{-1}})$ & $53.48$ & $1.0$\,dex \\
 & $\log(M_{\rm min}/M_\odot)$ & $6$ & $1.05$\,dex \\
 & LW amplitude $A_{\rm LW}^{31}$ & $0.35$ & $3.0$ \\
 & relic clumping $\log C_{\rm HII}$ & $0.60$ & $0.25$\,dex \\
 & relic temperature $\log(T_{25}/{\rm K})$ & $4$ & $0.3$\,dex \\
 & Ly$\alpha$ yield $\log f_{{\rm Ly}\alpha}$ & $-0.52$ & $0.3$\,dex \\
 & X-ray fraction $\log f_X$ & $-1.52$ & $0.5$\,dex \\
\end{tabular}
\end{ruledtabular}
\end{table}

\emph{Reference models.} Table~\ref{tab:params} lists the parameters of the
reference models. The flash relics have $R_{\rm HII}=1.47\cMpc$ and
$d_{\rm iso}=2.06\cMpc$. The flash raises $\tau$ from $0.060$ to $0.096$, moves
$z_{50}$ and $z_{90}$ from $6.83$ and $6.04$ to $6.90$ and $6.07$, and raises
$\Dthree$ from $1.64$ to $2.13\,\uK$.

\emph{Optical depth and history.} $\tau$ is computed from the mass-weighted
ionized fraction $\langle x_{\rm HII}(1+\delta)\rangle$, with helium singly
ionized alongside hydrogen and doubly ionized at $z=3$. A volume average would
under-count $\tau$ by 5--11\% during reionization, when ionized gas is
preferentially overdense. The history metrics $z_{50}$ and $z_{90}$ (the
redshifts at which half and $90\%$ of the volume is ionized) use the
volume-weighted fraction, as in the observational constraints.

\section{kSZ power spectrum}
\label{sec:ksz}

\emph{Fields.} We write the free-electron density in units of its fully
ionized mean, $e(\mathbf r)=n_e/\bar n_e=x_{\rm HII}(1+\delta)$, with helium
singly ionized wherever hydrogen is, and define the momentum field
$\mathbf q=e\,\mathbf v/c$. The kSZ temperature shift along the direction
$\hat{\mathbf n}$ is
\begin{equation}
  \frac{\Delta T}{T}(\hat{\mathbf n})=-\sigma_T\bar n_{e,0}\int d\chi\,(1+z)^2
  e^{-\tau(z)}\,\mathbf q\cdot\hat{\mathbf n},
\end{equation}
where $\bar n_{e,0}$ is the present-day mean density of hydrogen and singly
ionized helium electrons. In the Limber approximation only the component of
$\mathbf q$ transverse to the wavevector contributes, which gives
Eq.~(\ref{eq:limber}) of the Letter at $k=(\ell+1/2)/\chi$, where the factor
$1/2$ selects one of the two transverse components.

\emph{Transverse momentum power.} We neglect the connected four-point term and
take the velocities from linear theory, $\mathbf v(\mathbf k)/c=iA\,\delta_{\rm
lin}(\mathbf k)\,\mathbf k/k^2$ with $A=aHf/c$ and $f$ the linear growth rate.
The power of the transverse momentum is then~\cite{Ma:2001xr,Park:2013mv}
\begin{equation}
  P_{q_\perp}(k)=A^2\!\int\!\frac{d^3k'}{(2\pi)^3}\,(1-\mu^2)
  \left[P_{ee}(p)\,\frac{P_{\rm lin}(k')}{k'^2}
  -P_{e\delta}(p)\,\frac{P_{e\delta}(k')}{p^2}\right],
  \label{eq:pqlin}
\end{equation}
with $p=|\mathbf k-\mathbf k'|$, $\mu=\hat{\mathbf k}\cdot\hat{\mathbf k}'$,
$P_{\rm lin}$ the linear matter power spectrum and $P_{e\delta}$ the
cross-spectrum of $e$ with the linear density. The first term couples electron
fluctuations at $p$ to velocities at $k'$. The second is the
density--velocity cross term, which arises because the electrons trace the
same density field that sources the velocities, and lowers $\Dthree$ by
$\approx11\%$.

\emph{Velocities larger than the box.} We integrate over
$10^{-4}\leq k'\leq20\,{\rm Mpc^{-1}}$, far below the fundamental mode of the
box, $k_{\rm box}=2\pi/L=0.025\,{\rm Mpc^{-1}}$, so velocity modes larger than
the box are included~\cite{McQuinn:2005ce,Shaw:2011sy}. Restricting the
integral to $k'\geq k_{\rm box}$ lowers $\Dthree$ by $43\%$ without the flash
and by $42\%$ with it, so these modes change the two spectra by nearly the same
factor. The correction of Ref.~\cite{Park:2013mv}, used by
Ref.~\cite{Aggarwal:2026ogm}, is the same integral restricted to
$k'<k_{\rm box}$. With it, Ref.~\cite{Aggarwal:2026ogm} found the pkSZ at
$\ell=3000$ from $250\,h^{-1}\cMpc$ boxes to agree with a
$2\,h^{-1}\,{\rm cGpc}$ box to $\approx4.5\%$.

\emph{Electron spectra.} At each redshift node we measure $P_{ee}$,
$P_{\delta\delta}$ and $P_{e\delta}$ on the simulation grid in bins of
$0.1$\,dex in $k$. Equation~(\ref{eq:pqlin}) needs them at all wavenumbers. We
write $P_{e\delta}=b_eP_{\rm lin}$ with the measured ratio
$b_e=P_{e\delta}/P_{\delta\delta}$, in which the cosmic variance of the box
cancels. Near and beyond the box scale we model $e$ as a biased tracer with
white noise, $P_{ee}=S+b_0^2P_{\rm lin}$, and above the highest bin we continue
$P_{ee}$ as a power law.

\emph{Grid window.} \textsc{21cmFAST} smooths the gridded density twice, with
the top-hat filter of the initial conditions and with cloud-in-cell deposition.
At $0.5\cMpc$ cells this removes $3$, $8$ and $18\%$ of the density power at
$k=0.72$, $1.14$ and $1.8\,{\rm Mpc^{-1}}$. The ionized fraction is not
smoothed in this way, so we divide $P_{\delta\delta}$, $P_{e\delta}$ and the
density-correlated part of $P_{ee}$ by the window and discard bins where it
falls below $0.5$. This correction makes the pkSZ at $\ell=3000$ independent of
the cell size (Sec.~\ref{sec:conv}).

\emph{Redshift integral.} The spectra are measured on equal-time boxes at the
$\Delta\ln(1+z)=0.02$ nodes of each run between $z=5.5$ and $40$, and
Eq.~(\ref{eq:limber}) is integrated over these nodes, which avoids replicating
the box along the line of sight. The screening $\tau(z)$ follows from the
mass-weighted ionized fraction. Below $z=5.5$ helium is doubly ionized following
a $\tanh$ of width $0.5$ centred at $z=3$, which gives $\tau(z<5.5)=0.035$ and
multiplies every $D_\ell$ by $e^{-2\tau}\approx0.93$.

\emph{Components.} Following standard practice, the pkSZ is all kSZ from
$z>5.5$, i.e.\ Eq.~(\ref{eq:limber}) with $e=x_{\rm HII}(1+\delta)$. Writing
$e=(x_{\rm HII}-\bar x)(1+\delta)+\bar x(1+\delta)$, with $\bar x$ the
mass-weighted mean at each node, separates the fluctuations of the ionized
fraction (subscript $p$) from the mean ionized gas tracing the density
(subscript $h$),
\begin{equation}
  P_{pp}=P_{ee}-2\bar xP_{e\delta}+\bar x^2P_{\delta\delta},\qquad
  P_{hh}=\bar x^2P_{\delta\delta},
\end{equation}
plus their cross-spectrum. The fluctuation part gives $0.92$ and
$1.32\,\uK$ at $\ell=3000$ without and with the flash. The other two add
$0.72$ and $0.81\,\uK$, mostly from $z<8$, so the flash changes them by only
$0.08\,\uK$. The late-time kSZ ($z<5.5$) is not simulated. In the forecast it
is a template whose amplitude and tilt are marginalized
(Sec.~\ref{sec:forecast}).

\emph{SPT-3G limits.} The arrows in Fig.~\ref{fig:pksz} are $95\%$ upper limits
on the pkSZ power at $\ell=3000$, built as in Ref.~\cite{Aggarwal:2026ogm},
\begin{equation}
  D_{3000}^{\rm pkSZ}<D_{3000}^{\rm kSZ}+1.96\,\sigma\left(D_{3000}^{\rm kSZ}\right)-D_{3000}^{\rm hkSZ}.
\end{equation}
Here $D_{3000}^{\rm kSZ}$ is the total kSZ power measured by
Ref.~\cite{SPT-3G:2026mcr} for three foreground models, $3.96\pm0.82$ (Agora
templates), $2.5\pm1.1$ (free CIB and SZ) and $1.98\pm0.87\,\uK$ (free CIB,
power-law SZ), and $D_{3000}^{\rm hkSZ}=1.84\,\uK$ is the late-time kSZ
($0.85\,\uK$ in the low-hkSZ variant of the Agora case).

\section{Numerical convergence}
\label{sec:conv}

\emph{Cell size.} We test numerical convergence on a separate set of runs,
made with an earlier version of the flash model. These runs have the same
physics as the reference model but a shorter stellar lifetime
($t_{\rm life}=3$\,Myr) and hence smaller relics ($R_{\rm HII}=1.06\cMpc$), and
they reach a similar optical depth ($\tau=0.095$, of which $0.032$ at $z>12$).
They share the initial conditions of one $252\cMpc$ realization and differ only
in cell size, $1.0$, $0.75$ and $0.5\cMpc$. Between $1.0$ and $0.5\cMpc$,
$\tau$ changes by $2.1\%$ and the early optical depth $\tau(z>12)$ by
$0.9\%$. With the grid-window correction of Sec.~\ref{sec:ksz}, $\Dthree$
changes by less than $0.2\%$ and the power at $\ell=10^4$ by $3\%$, against
$1.5\%$ and $13\%$ without it. Coarsening the redshift step from
$\Delta\ln(1+z)=0.02$ to $0.04$ changes all of these by less than $0.6\%$. We
quote $\ell\le10^4$ in the Letter. The relics of the reference model are larger than those of the test
runs and are therefore better resolved at $0.5\cMpc$.

\emph{High-$\ell$ turnover.} In the same runs, the pkSZ spectrum turns over
where the relics become resolved. For randomly placed spheres of radius $R$ the
shot-noise contribution $D_\ell\propto\ell^2W^2(kR)$ peaks at $kR\approx2.1$,
that is, at $\ell\approx2.1\chi/R_{\rm HII}$. For $R_{\rm HII}=1.06\cMpc$ at
$z\approx21$--$25$ this predicts $\ell\approx21{,}600$--$22{,}300$. The
simulations give $19{,}600$ and $21{,}000$ at $0.75$ and $0.5\cMpc$, while the
Nyquist multipole moves from $46{,}000$ to $69{,}000$. The turnover is
therefore set by the relic size, not by the grid.

\emph{Realizations.} Over the four independent $252\cMpc$ realizations of the
reference model, $\Dthree$ scatters by $0.7\%$ (it is $2.5\%$ in $150\cMpc$
boxes), small enough for a Fisher forecast.

\section{Flash morphology and the high-$\ell$ spectrum}
\label{sec:morph}

\begin{figure*}
  \includegraphics[width=\textwidth]{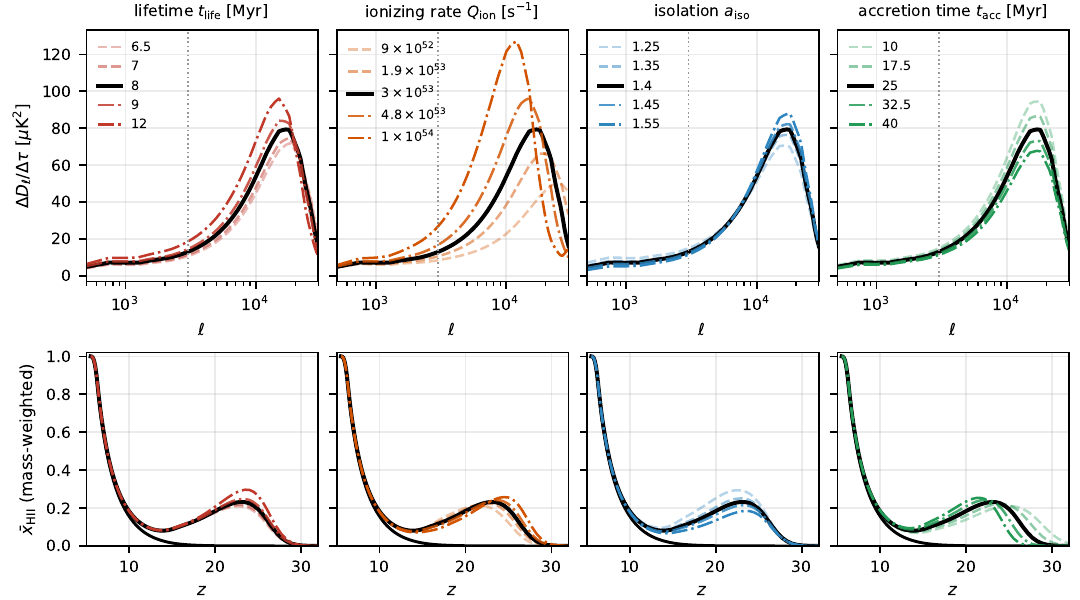}
  \caption{\textbf{How the flash morphology shapes its kSZ spectrum.} Top: the
  flash's pkSZ excess over the paired no-flash model, per unit of the optical
  depth it adds, when the lifetime (red), ionizing rate (orange), isolation
  factor (blue) or accretion time (green) is varied one at a time; dashed:
  below, dash-dotted: above the reference value; black: the reference flash.
  Dotted: $\ell=3000$. Bottom: mass-weighted ionization histories, with the
  no-flash model (thin black).}
  \label{fig:morphlog}
\end{figure*}

Figure~\ref{fig:morphlog} varies one property of the flash at a time around
the reference flash, in runs paired with the no-flash model (one realization,
the one with the widest set of variants). It shows the flash's kSZ excess per
unit of the optical depth it adds, $\Delta D_\ell/\Delta\tau$, which isolates
the morphology from the amount of ionization. The relic radius scales as
$R_{\rm HII}\propto(Q_{\rm ion}t_{\rm life})^{1/3}$ and the isolation distance
as $d_{\rm iso}=a_{\rm iso}R_{\rm HII}$.
\begin{itemize}
\item \emph{Relic size.} Raising $Q_{\rm ion}$ from $9\times10^{52}$ to
$10^{54}\,{\rm s^{-1}}$ ($R_{\rm HII}=1.0$ to $2.2\cMpc$) raises $\Delta D_\ell$ at $\ell=3000$ from $8$ to $27\,\uK$ per unit $\tau$ and
moves the high-$\ell$ peak from $\ell\approx2.6\times10^4$ to
$1.2\times10^4$; the lifetime acts the same way over its smaller range
of $R_{\rm HII}$ ($11$--$19\,\uK$ for $6.5$--$12$\,Myr).
\item \emph{Spacing.} Widening $d_{\rm iso}$ from $1.84$ to
$2.28\cMpc$ lowers $\Delta D_\ell$ at $\ell=3000$ from $15.0$ to
$11.5\,\uK$ per unit $\tau$ while leaving the power at
$\ell\approx10^4$ almost unchanged, as expected if isolation suppresses the
large-scale power ($S(k)<1$).
\item \emph{Timing.} $t_{\rm acc}$ sets timing: lengthening the accretion time from $10$ to $40$\,Myr moves the flash peak from $z\approx25$ to $21.6$ and lowers the $\Delta D_\ell$ at $\ell=3000$ from $15.0$ to $11.3\,\uK$ per unit $\tau$.
\end{itemize}

\section{Large-scale clustering of the flash relics}
\label{sec:bias}

The pkSZ at $\ell=3000$ is sourced by fluctuations of the ionized fraction on
comoving wavenumbers $k\approx\ell/\chi(z)$, i.e.\ $k\approx0.3\,{\rm
cMpc^{-1}}$ at the flash epoch ($\chi\approx11.2$\,cGpc at $z\approx23$). Here
we ask how the flash relics are clustered on these scales, and how much more
power they would carry if they traced their host halos.

\emph{Definitions.} Let $x(\mathbf r)$ be the ionized fraction of a cell,
$\bar x$ its volume average, $\delta x=x-\bar x$, and $\delta(\mathbf r)$ the
matter overdensity. For two fields $a$ and $b$ the power spectrum is defined by
$\langle\tilde a(\mathbf k)\tilde b^*(\mathbf k')\rangle=(2\pi)^3
\delta_D(\mathbf k-\mathbf k')P_{ab}(k)$; we measure $P_{xx}$, $P_{\delta\delta}$
and the cross-spectrum $P_{x\delta}$ on the $504^3$ grid in spherical shells in
$k$. If the ionized regions trace the matter linearly on large scales,
$\delta x=\bar x\,b_x\,\delta$, then $P_{x\delta}=\bar x\,b_x P_{\delta\delta}$.
We therefore define the bias of the ionized regions as
\begin{equation}
  b_x=\frac{1}{\bar x}\left\langle\frac{P_{x\delta}(k)}{P_{\delta\delta}(k)}\right\rangle,
  \label{eq:bx}
\end{equation}
where $\langle\cdot\rangle$ averages over $0.1\leq k\leq0.4\,{\rm cMpc^{-1}}$
weighted by the number of modes. Because the ratio uses the cross-spectrum
measured on the same density field, the cosmic variance of $\delta$ cancels and
uncorrelated (shot) power does not bias it. A negative $b_x$ means that the
ionized regions avoid overdense regions.

\emph{Power of isolated relics.} For $N$ relics of volume
$V_b=4\pi R_{\rm HII}^3/3$ in a volume $V$, with filling fraction
$\bar x_{\rm fl}=NV_b/V\ll1$,
\begin{equation}
  P_{xx}^{\rm fl}(k)\simeq\bar x_{\rm fl}\,V_b\,S(k)\,W^2(kR_{\rm HII}),
  \label{eq:shot}
\end{equation}
where $W$ is the Fourier transform of a single relic ($W\to1$ for
$kR_{\rm HII}\ll1$) and $S(k)$ is the structure factor of the source positions
($S=1$ for random placement, $S<1$ on scales larger than $d_{\rm iso}$ for
isolated sources). Integrated over $k$, Eq.~(\ref{eq:shot}) returns the
variance $\bar x_{\rm fl}$. Small relics spread it out to
$k\sim R_{\rm HII}^{-1}$ and leave only the plateau
$\bar x_{\rm fl}V_bS$ on the scales probed at $\ell=3000$
($kR_{\rm HII}\approx0.4$). Per unit $\tau$ this plateau grows with the relic
volume $V_b$ and falls with isolation.

\emph{Host-halo bias.} If the relics traced their hosts, $b_x$ would equal the
mean bias $\bar b_h$ of halos above the minimum host mass
$M_{\rm min}=10^6\,M_\odot$, which we compute from the Sheth--Tormen mass
function and halo bias~\cite{Sheth:1999mn}, weighted by halo number. Weighting
by collapsed mass instead raises $\bar b_h$ by $\lesssim10\%$.

\begin{figure}[h]
  \includegraphics[width=0.6\textwidth]{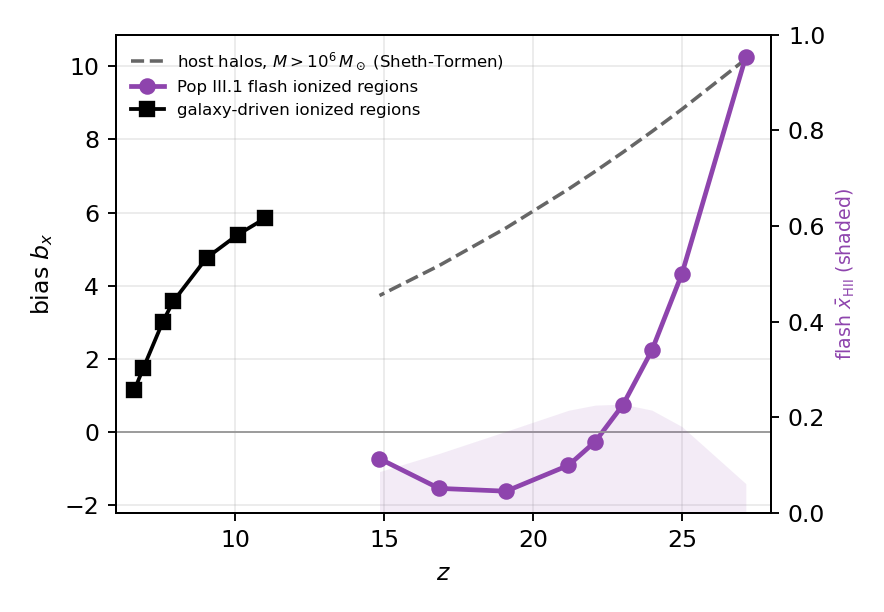}
  \caption{\textbf{The flash relics lose the bias of their host halos.} Bias
  $b_x$ [Eq.~(\ref{eq:bx})], averaged over $0.1\leq k\leq0.4\,{\rm
  cMpc^{-1}}$, of the flash ionized regions in the reference flash run
  (purple; realization 1), of halos above $M_{\rm min}=10^6\,M_\odot$
  (dashed) and of the galaxy-driven ionized regions in the no-flash run
  (black). Shaded (right axis): the volume-averaged flash ionized fraction.}
  \label{fig:bias}
\end{figure}

\emph{Results.} Figure~\ref{fig:bias} shows $b_x$ for the reference flash. At
ignition the relics trace their hosts: at $z=27$, $b_x=10.3$ against
$\bar b_h=10.2$. By the flash peak they have lost this clustering: at $z=23$,
$b_x=0.6$--$0.7$ across the four realizations against $\bar b_h=7.7$.
Afterwards they avoid overdense regions ($b_x\approx-1.6$ at
$z\approx17$--$19$). From the flash peak on, their power at these $k$ is
$0.4$--$0.5$ of that of randomly placed relics [Eq.~(\ref{eq:shot}) with
$S=1$], as expected for isolated sources ($S<1$). For
comparison, galaxy-driven ionized regions in the no-flash run have $b_x=5.4$,
$3.6$ and $1.8$ at $z=10$, $8$ and $7$.

\emph{Power of a host-biased flash.} If the relics had kept the host bias,
their ionized fraction would carry the additional large-scale term
$\delta x=\bar x\,\bar b_h\,\delta$, and their power would be
\begin{equation}
  P_{xx}^{\rm host}(k)\simeq P_{xx}(k)+\bar b_h^2\,\bar x^2P_{\delta\delta}(k).
  \label{eq:hostpower}
\end{equation}
We evaluate Eq.~(\ref{eq:hostpower}) in the bin $k=0.29\,{\rm cMpc^{-1}}$,
the one closest to the Limber wavenumber $k=\ell/\chi$ of $\ell=3000$ at these
redshifts, and find $P_{xx}^{\rm host}/P_{xx}=3.8$, $9.6$ and $10.5$ at $z=25$,
$23$ and $21$. Near the flash peak ($z=21$--$23$) the host term
$\bar b_h^2\bar x^2P_{\delta\delta}$ is $9$ times the measured relic power
$P_{xx}$ at $\ell=3000$, $3$ times at $\ell=6000$ and $1.2$ times at
$\ell=10^4$. It therefore dominates the relic power at all $\ell\lesssim10^4$.
A host-biased flash would therefore boost the
spectrum at the multipoles where the pkSZ is measured rather than add power at
high $\ell$. 

\emph{Comparison with Ref.~\cite{Aggarwal:2026ogm}.} Ref.~\cite{Aggarwal:2026ogm}
seeded Pop~III.1 bubbles on density peaks with an isolation criterion and
found that a flash with $\Delta\tau\approx0.03$ adds
$\approx1.9\,\uK$ at $\ell=3000$ in their fiducial case, which
evades only the most conservative SPT-3G limit. Randomly placed sources add
much less. Our self-consistent flash uses structurally the same biased,
exclusion-based seeding but costs $\approx5$ times less per unit
$\tau$, and its excess rises with $\ell$ instead of boosting the spectrum near
$\ell=3000$. The host-bias term above, a factor of 4--10 in the relic
power, is large enough to account for the difference.

\section{Forecast}
\label{sec:forecast}

We adopt the Gaussian likelihood and parameter set of
Ref.~\cite{Raghunathan:2023yfe} (hereafter R\&O), a published Fisher forecast of
the kSZ power spectrum for SO, validate our implementation against its
published results, and then make three changes that serve the question of
this Letter. We leave $\tau$ free, describe the pkSZ by a free amplitude in each
multipole band and add a simulated reionization model with external data. We
also adopt SO's official noise curves and treat the ILC residual as known.

\subsection{Validation}

We reproduce R\&O's headline result, the detection significance of the
total kSZ power $D_\ell=A_{\rm kSZ}\times3\,\uK$ for SO goal (their Table~4),
with our own implementation of their likelihood and their public inputs. This
tests the noise treatment, the covariance and the marginalization over
cosmology, but not our reionization models, which do not enter.
R\&O quote two foreground treatments: the minimum-variance (MV) ILC, and a
cross-ILC that cross-correlates a tSZ-deprojected map with a CIB-deprojected
one, which removes their power from the kSZ spectrum at the cost of higher
noise. Table~\ref{tab:ro_valid} shows agreement to $\lesssim5\%$ ($8\%$ in one case,
where R\&O note a degeneracy with the radio residuals that we treat more
simply).

\begin{table}[h]
\caption{Total-kSZ S/N $=1/\sigma(A_{\rm kSZ})$ for SO goal ($f_{\rm sky}=0.4$),
as a function of the maximum multipole at which the kSZ is used.}
\label{tab:ro_valid}
\begin{ruledtabular}
\begin{tabular}{lcccc}
 & $\ell_{\max}^{\rm kSZ}=3000$ & $3500$ & $4000$ & $4500$ \\
\hline
R\&O Table 4, MV-ILC & 16.1 & 27.1 & 35.5 & 32.8 \\
this work, MV-ILC & 15.5 & 25.8 & 35.1 & 35.5 \\
R\&O Table 4, cross-ILC & 5.5 & 8.5 & 11.9 & 14.9 \\
this work, cross-ILC & 5.4 & 8.2 & 11.5 & 14.6 \\
\end{tabular}
\end{ruledtabular}
\end{table}

\subsection{Data model and compressed likelihood}
\label{sec:datamodel}

The CMB data are the TT, EE and TE power spectra of the component-separated SO
map at $\ell=30$--$6900$ and the lensing-reconstruction spectrum
$C_L^{\kappa\kappa}$ at $30\leq L\leq1000$. At each multipole the covariance of
the three primary spectra is
\begin{equation}
  \Sigma_\ell=\frac{1}{(2\ell+1)f_{\rm sky}}
  \begin{pmatrix}
   2\tilde T^2 & 2\,(C^{TE})^2 & 2\,\tilde T\,C^{TE}\\
   2\,(C^{TE})^2 & 2\,\tilde E^2 & 2\,\tilde E\,C^{TE}\\
   2\,\tilde T\,C^{TE} & 2\,\tilde E\,C^{TE} & (C^{TE})^2+\tilde T\tilde E
  \end{pmatrix},
  \label{eq:ro_cov}
\end{equation}
with $\tilde T=C^{TT}+N^T$ and $\tilde E=C^{EE}+N^P$. For SO goal, $N^T$ is the SO
collaboration's internal-linear-combination (ILC) noise curve, which contains
the instrumental noise and the CIB, tSZ and radio power the ILC leaves
(``Deproj-0''~\cite{SimonsObservatory:2018koc}), and $N^P$ and the lensing noise are SO's
published curves. We treat the foreground residual as known: its power enters
the variance, but no foreground amplitude is marginalized. The lensing spectrum
is treated as independent of the primary spectra. The temperature signal is
\begin{equation}
  C_\ell^{TT}=C_\ell^{\rm CMB}(\boldsymbol\theta_\Lambda,\tau)
  +C_\ell^{\rm pkSZ}
  +\frac{2\pi}{\ell(\ell+1)}\,A_h\,D_{3000}^{h}\,T(\ell)\left(\frac{\ell}{3000}\right)^{\alpha_h},
\end{equation}
with $C^{\rm CMB}$ the lensed Planck 2018 spectra and $C^{\rm pkSZ}$ our
simulated pkSZ. The last term is the late-time ($z<5.5$) kSZ: $T(\ell)$ is the
template of Ref.~\cite{Battaglia:2010tm} normalized at $\ell=3000$, and
$D^{h}_{3000}=1.65\,\uK\,(\sigma_8/0.8)^{4.46}=1.75\,\uK$ for our
$\sigma_8$~\cite{SPT-3G:2026mcr,Calabrese:2014gwa}. Its amplitude $A_h$ has a $13\%$
Gaussian prior, the $1.5\pm0.2\,\uK$ of Ref.~\cite{SimonsObservatory:2018koc}, and the tilt
$\alpha_h$ a prior of width $0.3$.

\emph{Compression.} We describe the pkSZ by a free amplitude in each multipole
band. The band edges are $\ell=30$, $1500$, $2500$, $3500$, $4500$, $5500$,
$6500$ and $6900$, giving bands centred at $\ell=2000$--$6000$ with
$\Delta\ell=1000$ and two edge bands. In each band the fiducial spectrum is
raised or lowered by a constant in $D_\ell$, so that the band power $D_b$, the
mean of $D_\ell^{\rm pkSZ}$ over the band, is free. All seven bands are free,
so that the pkSZ outside the bands we use is marginalized rather than fixed. We
compute the Fisher matrix of the band powers, $\Lambda$CDM ($A_s$, $n_s$,
$\omega_b$, $\omega_c$, $\theta_*$), $\tau$ and the two late-time kSZ
parameters from the TT, EE and TE spectra with the covariance of
Eq.~(\ref{eq:ro_cov}) and from the lensing spectrum, with priors only on the
late-time kSZ. Inverting it and keeping the rows and columns of the band powers
and $\tau$ gives the joint Gaussian covariance $\mathsf C$ of the band powers
and of $\tauthirty$, the optical depth from the same data without large-scale
polarization. Here $\tau$ enters the primary spectra only through
$\ell\geq30$, and the lensing spectrum, which does not depend on $\tau$, breaks
its degeneracy with $A_s$. The pair $(\boldsymbol D,\tauthirty)$ with
covariance $\mathsf C$ is the CMB likelihood of the Letter, which we combine
with the reionization model below. Every data combination of
Fig.~\ref{fig:tauplane} and Table~\ref{tab:sigtau} uses a subset of it (the
bands at $\ell=3000$--$6000$, $\tauthirty$, or both), so adding data can only
tighten the bounds.

The diagonal of this covariance gives the bandpower errors of
Fig.~\ref{fig:pksz}, $\sigma(D_b)=0.35$, $0.25$,
$0.26$, $0.28$ and $0.32\,\uK$ at
$\ell=2000$--$6000$. Above $\ell=3000$ they are dominated by the late-time kSZ
prior ($0.13\times1.75\,\uK=0.23\,\uK$), which also makes neighbouring bands
94--96\% correlated, so the shape information resides in band
differences. The bands and $\tauthirty$ are almost uncorrelated
($\lesssim8\%$). On its own, $\sigma(\tauthirty)=0.0086$,
which comes from the damping-tail amplitude $A_se^{-2\tau}$ with $A_s$ fixed
by lensing; Planck temperature, polarization and lensing at $\ell\geq30$ give
$0.016$~\cite{Giare:2023ejv}. We impose no external $\tau$.

\subsection{Reionization model and parameter forecast}
\label{sec:reionfisher}

\emph{Observables.} For a set of reionization parameters $\mathbf p$ the
simulations predict three groups of observables: the band powers
$D_b(\mathbf p)$ at $\ell=3000$--$6000$, averaged over each band; the optical depth $\tau(\mathbf p)=\tau_{\rm
late}+\tau_{\rm early}$, the mass-weighted optical depth split at $z=12$,
which is what $\tauthirty$ measures; and the forest observables
$\mathbf x(\mathbf p)$, the volume-weighted neutral fraction at the redshifts
of the four damping-wing measurements and the redshift $z_{5\%}$ at which it
falls to $5\%$.

\emph{Likelihood.} The data vector $\mathbf d$ collects the measured band
powers and $\tauthirty$ ($\mathbf d_{\rm CMB}$) and the forest data
($\mathbf d_x$, with errors $\sigma_i$). The log-likelihood is
\begin{equation}
  -2\ln\mathcal L=\left[\mathbf d_{\rm CMB}-\mathbf m(\mathbf p)\right]^{\rm T}
  \mathsf C^{-1}\left[\mathbf d_{\rm CMB}-\mathbf m(\mathbf p)\right]
  +\sum_i\frac{[d_{x,i}-x_i(\mathbf p)]^2}{\sigma_i^2}
  +\sum_a\frac{(p_a-\bar p_a)^2}{\sigma_{{\rm prior},a}^2},
  \label{eq:like}
\end{equation}
with $\mathbf m=(D_b,\tau)$ and $\bar p_a$ the fiducial values. Its Fisher
matrix is
\begin{equation}
  \mathcal F_{ab}=\partial_a\mathbf m^{\rm T}\mathsf C^{-1}\partial_b\mathbf m
  +\sum_i\frac{\partial_ax_i\,\partial_bx_i}{\sigma_i^2}
  +\frac{\delta_{ab}}{\sigma_{{\rm prior},a}^2}.
  \label{eq:fisher2}
\end{equation}
A data combination keeps only the corresponding rows of $\mathbf m$ and
$\mathsf C$, or drops the forest term. The two partial optical depths are
derived quantities, so their covariance follows by propagation,
\begin{equation}
  {\rm Cov}(\tau_{\rm late},\tau_{\rm early})=\mathsf E\,\mathcal F^{-1}\mathsf E^{\rm T},
  \qquad \mathsf E_{ia}=\frac{\partial\tau_i}{\partial p_a},
  \label{eq:taucov}
\end{equation}
which is what Fig.~\ref{fig:tauplane} and Tables~\ref{tab:sigtau},
\ref{tab:surveys} and~\ref{tab:rob} show. All derivatives are central
differences, $\partial_a\mathbf m=[\mathbf m(\bar p_a+h_a)-\mathbf m(\bar
p_a-h_a)]/2h_a$, around the fiducial flash, with each of the $\pm$ runs
averaged over the four realizations (one-sided for $M_{\rm min}$).

\emph{Parameters.} Table~\ref{tab:params} lists the 13 parameters, their
fiducial values and their priors. The steps $h_a$ are those of the simulation
suite of Ref.~\cite{Kern:inprep}. The three galaxy parameters carry
the UV-LF priors of Ref.~\cite{Park:2018ljd}, and $F_{\rm esc,10}$ is free
(Sec.~\ref{sec:uvlf}). The turnover mass, $\alpha_{\rm esc}$, $t_\star$ and the
minihalo parameters are fixed. The ten flash parameters carry the priors of
Ref.~\cite{Kern:inprep}. The flash amplitude, set by the occupation fraction
of the source halos, is fixed at unity. Together with the ten parameters a
free amplitude would open a direction (a stronger flash of shorter-lived,
dimmer sources) that the data cannot close.

The flash parameters are defined in Ref.~\cite{Kern:inprep}. The first five
set the sources and their relics: the isolation distance
$d_{\rm iso}=a_{\rm iso}R_{\rm HII}$, the ionizing lifetime and photon rate
that fix $R_{\rm HII}$ through Eq.~(\ref{eq:RHII}), the delay
$t_{\rm acc}$ between halo collapse and ignition (the dark-star phase), and the
minimum host mass $M_{\rm min}$. $A_{\rm LW}^{31}$ is the amplitude of the
Lyman--Werner feedback that raises the minimum host mass as the radiation
background builds up, with the same form as the minihalo feedback of
\textsc{21cmFAST}. The last four set how the relics evolve and what else the
sources emit: the sub-cell clumping factor in the relic recombination rate,
the photoheating-equilibrium temperature of the relics,
$T(z)=T_{25}(1+25)/(1+z)$, which sets their recombination coefficient, the
Lyman-band photon yield $f_{{\rm Ly}\alpha}=Q_{{\rm Ly}\alpha}/Q_{\rm ion}$,
and the soft X-ray fraction $f_X$ of the bolometric luminosity of the
accreting black holes left behind by the sources.

\emph{Forest.} The four damping-wing measurements~\cite{Davies:2018yfp,Wang:2020zae,Yang:2020J1007}
enter with their symmetrized $68\%$ errors. The end of reionization enters as
$z_{5\%}=5.55\pm0.25$, a Gaussian that corresponds to
$x_{\rm HI}\lesssim5\%$ at $5.3<z<5.8$~\cite{Cain:2025usc,Bosman:2021oom}
(Sec.~\ref{sec:ppc}). In a Fisher forecast only the widths matter.

\subsection{Current data}
\label{sec:current}
The current-data result of the Letter applies the same model to existing
measurements. We linearize the observables around the fiducial flash,
$\mathbf m(\mathbf p)\simeq\mathbf m(\bar{\mathbf p})+\partial\mathbf m\,
\delta\mathbf p$, so that Eq.~(\ref{eq:like}) is quadratic in
$\delta\mathbf p=\mathbf p-\bar{\mathbf p}$. Its maximum is at
\begin{equation}
  \delta\hat{\mathbf p}=\mathcal F^{-1}\mathbf b,\qquad
  b_a=\sum_i\frac{\partial_ax_i\,[d_{x,i}-x_i(\bar{\mathbf p})]}{\sigma_i^2}
  +\frac{\partial_a\tau\,[\tauthirty-\tau(\bar{\mathbf p})]}{\sigma^2(\tauthirty)},
\end{equation}
with $\mathcal F$ from Eq.~(\ref{eq:fisher2}) restricted to these data, and
the partial optical depths are $\hat\tau_i=\tau_i(\bar{\mathbf p})+\mathsf
E_{ia}\delta\hat p_a$ with the covariance of Eq.~(\ref{eq:taucov}). The data
are the four damping wings at their measured values, $z_{5\%}=5.55\pm0.25$,
and the Planck optical depth without large-scale polarization,
$\tauthirty=0.078\pm0.016$~\cite{Giare:2023ejv}, with the same 13 parameters
and priors as the forecast. This gives $\tau_{\rm late}=0.058\pm0.002$ and
$\tau_{\rm early}=0.023\pm0.015$.

To add the SPT-3G measurement~\cite{SPT-3G:2026mcr} we compare its total kSZ
power at $\ell=3000$ with the model band power at $\ell=2500$--$3500$ plus the
late-time kSZ, $D_{3000}(\mathbf p)+D^h_{3000}$, and add the late-time
amplitude uncertainty ($13\%$ of $1.75\,\uK$) in quadrature to the SPT error.
For the three foreground models of Sec.~\ref{sec:ksz} this gives
$\tau_{\rm early}=0.026$, $0.019$ and $0.016$ (Agora, free CIB and SZ, free
CIB with power-law SZ), each with $\sigma=0.014$. Unlike the forecast, these
values depend on the absolute amplitude of the simulated pkSZ, since the
measured power is compared with the predicted one.

\subsection{Other surveys}
\label{sec:surveys}

Table~\ref{tab:surveys} repeats the forecast for three other surveys with the
same likelihood, foreground treatment and late-time kSZ model.
\emph{SO enhanced} is the current SO configuration~\cite{SimonsObservatory:2025wwn}:
its depths, beams, atmospheric noise and $f_{\rm sky}=0.4$ are published, but
its ILC and lensing noise are not, so we use curves derived from the published
configuration with a foreground model built from ACT DR6 fits; the same
pipeline reproduces SO goal's official curves to $\lesssim7\%$ over the pkSZ
range. \emph{SPT-3G} uses the ILC residuals and lensing noise of R\&O on
$1500\,{\rm deg}^2$; its own $\tauthirty$ is combined with Planck's
($\sigma=0.016$~\cite{Giare:2023ejv}) as an independent measurement, since the two
skies barely overlap. \emph{CMB-HD} uses the collaboration's noise and
foreground curves~\cite{Sehgal:2019ewc,MacInnis:2023vif} on $f_{\rm sky}=0.6$. The late-time kSZ
prior sets a floor of $\approx0.23\,\uK$ on every bandpower, which is why the
pkSZ improves only slowly with depth; the gains come from $\tauthirty$.

\begin{table}[h]
\caption{Other surveys: $\sigma(\tauthirty)$ from the survey alone (for SPT-3G,
then combined with Planck), the bandpower errors at $\ell=3000$--$6000$, and
$\sigma(\tau_{\rm early})$ around the fiducial flash with the ten flash
parameters free and the UV-LF priors.}
\label{tab:surveys}
\begin{ruledtabular}
\begin{tabular}{lcccccc}
survey & $\sigma(\tauthirty)$ & $\sigma(D_b)$ [$\uK$] & pkSZ & pkSZ + forest & forest + $\tauthirty$ & all \\
\hline
SO goal     & 0.0086 & 0.25--0.32 & 0.022 & 0.015 & 0.0087 & 0.0077 \\
SO enhanced & 0.0079 & 0.24--0.28 & 0.020 & 0.014 & 0.0081 & 0.0072 \\
SPT-3G      & 0.025  & 0.28--0.38 & 0.025 & 0.016 & 0.013  & 0.010 \\
CMB-HD      & 0.0056 & 0.22--0.23 & 0.017 & 0.013 & 0.0060 & 0.0054 \\
\end{tabular}
\end{ruledtabular}
\end{table}

\subsection{Robustness}

Table~\ref{tab:rob} varies the assumptions of the SO goal forecast one at a
time. The combinations that include $\tauthirty$ change by at most
$10\%$ for any choice except the widths of the flash priors. The
late-time kSZ amplitude matters most for the pkSZ with the forest: a $25\%$
prior, the range spanned by the current $\sigma_8$ tension, loosens it by
$33\%$, and a free amplitude doubles it, while the bound with
$\tauthirty$ moves by $\lesssim8\%$. The tilt of the late-time kSZ and
the foreground amplitude change nothing material. The end-of-reionization prior
tightens $\tau_{\rm late}$ by 40\% but barely moves $\tau_{\rm early}$.
Halving or doubling all ten flash prior widths changes the full combination by
$-12$ and $+10\%$, and the pkSZ alone by a factor of
$\approx1.7$: without $\tauthirty$, the flash bounds depend on how
well its morphology is known. Freeing $\alpha_\star$ is discussed in Sec.~\ref{sec:uvlf}. The
$\ell=6000$ band does not drive the results. Without it,
$\sigma(\tau_{\rm early})$ is 0.023 for the pkSZ alone (four bands: 0.022),
0.016 with the forest (0.015), and 0.0077 with the forest and $\tauthirty$
(0.0077).

\begin{table}[h]
\caption{Robustness of the SO goal forecast: $\sigma(\tau_{\rm early})$ around
the fiducial flash (ten flash parameters free, UV-LF priors), and
$\sigma(\tau_{\rm late})$ for all data.}
\label{tab:rob}
\begin{ruledtabular}
\begin{tabular}{lccccc}
variant & pkSZ & pkSZ + forest & forest + $\tauthirty$ & all & $\sigma(\tau_{\rm late})$, all \\
\hline
baseline                                   & 0.022 & 0.015 & 0.0087 & 0.0077 & 0.0021 \\
late-time kSZ amplitude prior $25\%$       & 0.023 & 0.020 & 0.0087 & 0.0080 & 0.0021 \\
late-time kSZ amplitude free               & 0.031 & 0.029 & 0.0087 & 0.0083 & 0.0022 \\
late-time kSZ tilt fixed / free            & 0.022 & 0.015 & 0.0087 & 0.0077 & 0.0021 \\
CIB+tSZ amplitude, $10\%$ prior            & 0.023 & 0.015 & 0.0087 & 0.0077 & 0.0021 \\
forest: damping wings only                 & 0.022 & 0.015 & 0.0090 & 0.0081 & 0.0034 \\
forest: all compiled points & 0.022 & 0.015 & 0.0086 & 0.0075 & 0.0015 \\
flash prior widths $\times0.5$             & 0.013 & 0.011 & 0.0080 & 0.0068 & 0.0017 \\
flash prior widths $\times2$               & 0.036 & 0.020 & 0.0093 & 0.0085 & 0.0031 \\
\end{tabular}
\end{ruledtabular}
\end{table}

\section{Role of the UV luminosity functions}
\label{sec:uvlf}
The galaxy model of the forecast has three free parameters: the normalization
$F_{\star,10}$ and slope $\alpha_\star$ of the star-formation efficiency with
halo mass, and the escape-fraction normalization $F_{\rm esc,10}$
($\alpha_{\rm esc}$, the turnover mass and the star-formation timescale are
fixed). The UV LFs enter through the LF-only constraints of
Ref.~\cite{Park:2018ljd}, which uses the same \textsc{21cmFAST} galaxy
parametrization, as Gaussian likelihoods of width $0.30$\,dex on $F_{\star,10}$
and $0.07$ on $\alpha_\star$. The LFs measure the non-ionizing UV light of
galaxies and are insensitive to the escape fraction, so $F_{\rm esc,10}$ is
left free.

The slope $\alpha_\star$ sets how steeply star formation rises with halo mass,
and hence how drawn-out galaxy-driven reionization is. A shallower slope adds
pkSZ power at all scales while barely changing $\tau_{\rm late}$
($dD_{3000}/d\alpha_\star\approx-0.4\,\uK$ against $d\tau_{\rm
late}/d\alpha_\star\approx0.003$). This is the degeneracy between the
timing and the duration of reionization that limits the pkSZ in full
\textsc{21cmFAST} inferences, where the LFs also break it~\cite{Nikolic:2023wea}.
Without the calibration, the duration of the late history absorbs part of the
spectral shape that separates the flash from galaxies.

\begin{figure}[h]
  \includegraphics[width=\textwidth]{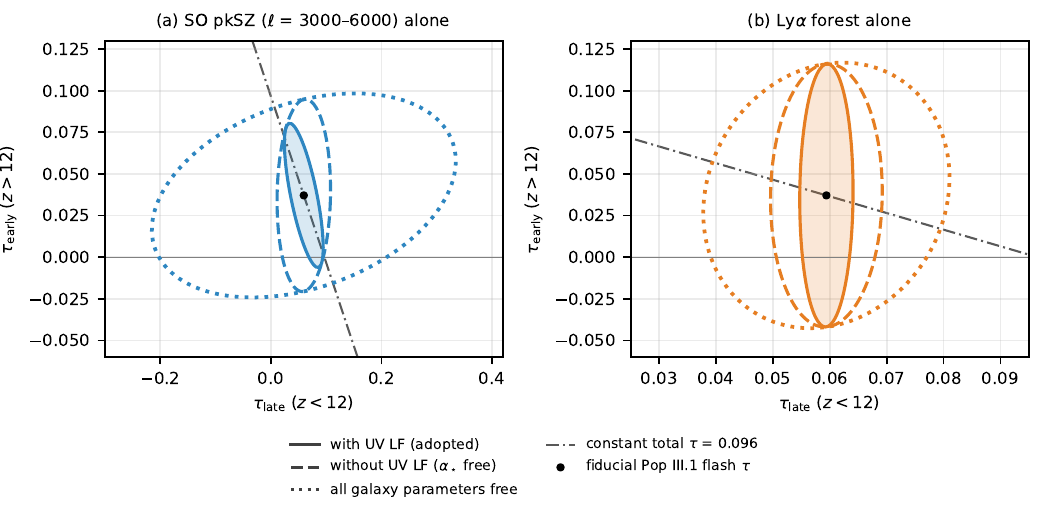}
  \caption{\label{fig:uvlf_contours} The pkSZ alone (a) and the Ly$\alpha$ forest
  alone (b) for SO goal around the fiducial flash, with the ten flash parameters
  free, for three treatments of the galaxy model: calibrated on the UV LFs (solid,
  filled; adopted), with the star-formation slope $\alpha_\star$ free (dashed),
  and with all three galaxy parameters free (dotted). Ellipses span the
  marginalized $\pm2\sigma$ errors. Dash-dotted: constant total $\tau$ through the
  fiducial. Panel (a) is zoomed out to show the all-free case.}
\end{figure}

Figure~\ref{fig:uvlf_contours} compares the pkSZ alone and the forest alone with
this calibration, with $\alpha_\star$ free (keeping a $1$\,dex width on
$F_{\star,10}$, which only regularizes the direction along which it trades off
against $F_{\rm esc,10}$), and with all three galaxy parameters free. The
ionization history depends on $F_{\star,10}$ and $F_{\rm esc,10}$ almost only
through their product (their responses agree to 2--4\%); they differ
only through the X-ray and Ly$\alpha$ heating that follows the star formation, so
with both free the pkSZ loses $\tau_{\rm late}$ almost entirely
($\sigma=0.14$). Freeing $\alpha_\star$ alone widens the pkSZ ellipse
about the line of constant total $\tau$ and doubles the forest's error on
$\tau_{\rm late}$, from 0.0023 to 0.0049: the forest fixes
the end of the late history but not how drawn-out it is.

Of the combinations in Table~\ref{tab:sigtau}, the pkSZ with the forest is
the most sensitive to the calibration: freeing $\alpha_\star$ raises its
$\sigma(\tau_{\rm early})$ from 0.015 to 0.021. With $\tauthirty$ the loss is
20\%.

\section{Ly\texorpdfstring{$\alpha$}{alpha}-forest data}
\label{sec:ppc}

\begin{figure}[h]
  \includegraphics[width=\textwidth]{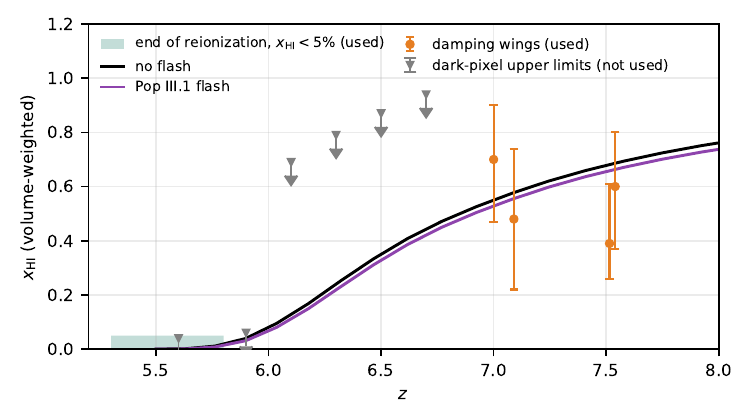}
  \caption{\textbf{The reference histories and the Ly$\alpha$-forest data.}
  Volume-weighted neutral fraction of the no-flash (black) and flash (purple)
  reference models, with the damping-wing measurements and the end-of-reionization
  window used in the forecast, and the dark-pixel upper limits (not used).}
  \label{fig:ppc}
\end{figure}

The forecast uses two forest observables (Sec.~\ref{sec:forecast}): the
quasar damping-wing measurements of the neutral fraction at
$z=7.0$--$7.5$~\cite{Davies:2018yfp,Wang:2020zae,Yang:2020J1007}, and the end of
reionization, $x_{\rm HI}\lesssim5\%$ at $5.3<z<5.8$~\cite{Cain:2025usc,Bosman:2021oom}.
Figure~\ref{fig:ppc} compares them with the reference histories. The damping
wings give $\chi^2=3.2$ (flash) and $3.6$ (no flash) for four
points. Both models reach $x_{\rm HI}=5\%$ at $z=5.95$ and
$5.92$, $1.6\sigma$ above the centre of the end window, and
both satisfy the dark-pixel upper limits~\cite{McGreer:2014qwa,Jin:2022ngn}, which we
show but do not use. Because a Fisher forecast centres each constraint on the
fiducial, this offset does not affect the errors; it would matter for a fit to
the data. Using all the measurements in our compilation, or
dropping the end prior, changes $\sigma(\tau_{\rm early})$ by
$\lesssim5\%$ (Table~\ref{tab:rob}).

\end{document}